\documentstyle[epsfig]{elsart}

\begin{document}

\begin{frontmatter}
\title{A combined analysis technique for the \\
search for fast magnetic monopoles with the MACRO detector}
 
\begin{center}
{\rm The MACRO Collaboration} \\

\nobreak\bigskip\nobreak
\pretolerance=10000
M.~Ambrosio$^{12}$, 
R.~Antolini$^{7}$, 
G.~Auriemma$^{14,a}$, 
D.~Bakari$^{2,17}$, 
A.~Baldini$^{13}$, 
G.~C.~Barbarino$^{12}$, 
B.~C.~Barish$^{4}$, 
G.~Battistoni$^{6,b}$, 
Y.~Becherini$^{2}$,
R.~Bellotti$^{1}$
% ...
\\
\vspace{0.3 cm}
\footnotesize
1. Dipartimento di Fisica dell'Universit\`a  di Bari and INFN, 70126 Bari, Italy \\
2. Dipartimento di Fisica dell'Universit\`a  di Bologna and INFN, 40126 Bologna, Italy \\
3. Physics Department, Boston University, Boston, MA 02215, USA \\
4. California Institute of Technology, Pasadena, CA 91125, USA \\
5. Department of Physics, Drexel University, Philadelphia, PA 19104, USA \\
6. Laboratori Nazionali di Frascati dell'INFN, 00044 Frascati (Roma), Italy \\
7. Laboratori Nazionali del Gran Sasso dell'INFN, 67010 Assergi (L'Aquila), Italy \\
8. Depts. of Physics and of Astronomy, Indiana University, Bloomington, IN 47405, USA \\
9. Dipartimento di Fisica dell'Universit\`a  dell'Aquila and INFN, 67100 L'Aquila, Italy\\
10. Dipartimento di Fisica dell'Universit\`a  di Lecce and INFN, 73100 Lecce, Italy \\
11. Department of Physics, University of Michigan, Ann Arbor, MI 48109, USA \\
12. Dipartimento di Fisica dell'Universit\`a  di Napoli and INFN, 80125 Napoli, Italy \\
13. Dipartimento di Fisica dell'Universit\`a  di Pisa and INFN, 56010 Pisa, Italy \\
14. Dipartimento di Fisica dell'Universit\`a  di Roma "La Sapienza" and INFN, 00185 Roma, Italy \\
15. Physics Department, Texas A\&M University, College Station, TX 77843, USA \\
16. Dipartimento di Fisica Sperimentale dell'Universit\`a  di Torino and INFN, 10125 Torino, Italy \\
17. L.P.T.P, Faculty of Sciences, University Mohamed I, B.P. 524 Oujda, Morocco \\
18. Dipartimento di Fisica dell'Universit\`a  di Roma Tre and INFN Sezione Roma Tre, 00146 Roma, Italy \\
$a$ Also Universit\`a  della Basilicata, 85100 Potenza, Italy \\
$b$ Also INFN Milano, 20133 Milano, Italy \\
$c$ Also Istituto TESRE/CNR, 40129 Bologna, Italy \\
$d$ Also Universit\`a  di Trieste and INFN, 34100 Trieste, Italy \\
$e$ Also Dipartimento di Energetica, Universit\`a  di Roma, 00185 Roma, Italy \\
$f$ Also Institute for Nuclear Research, Russian Academy of Science, 117312 Moscow, Russia \\
$g$ Also Institute for Space Sciences, 76900 Bucharest, Romania \\
$h$ Macalester College, Dept. of Physics and Astr., St. Paul, MN 55105 \\
$i$ Also Department of Physics, Pennsylvania State University, University Park, PA 16801, USA \\
$l $Also Dipartimento di Fisica dell'Universit\`a  della Calabria, Rende (Cosenza), Italy \\
$m$ Also Department of Physics, James Madison University, Harrisonburg, VA 22807, USA \\
$n$ Also RPD, PINSTECH, P.O. Nilore, Islamabad, Pakistan \\
$o$ Also Dipartimento di Scienze e Tecnologie Avanzate, Universit\`a  del Piemonte Orientale, Alessandria, Italy \\
$p$ Also U. Minn. Duluth Physics Dept., Duluth, MN 55812 \\
$q$ Also Dept. of Physics, MIT, Cambridge, MA 02139 \\
$r$ Also Intervideo Inc., Torrance CA 90505 USA \\
$s$ Also Resonance Photonics, Markham, Ontario, Canada\\

\end{center}

\begin{abstract}

We describe a search method for fast moving ($\beta > 5 \times 10^{-3}$)
magnetic monopoles using simultaneously 
the scintillator, streamer tube and track-etch subdetectors
of the MACRO apparatus.
The first two subdetectors are used primarily for the identification
of candidates while the track-etch one is used as the final tool for their
rejection or confirmation.
Using this technique, a first sample of more than two years of data has been
analyzed without any evidence of a magnetic monopole.
We set a $90 \%$ CL upper limit to the local monopole flux of 
$1.5 \times 10^{-15} \, cm^{-2} s^{-1} sr^{-1}$ in 
the velocity range $5 \times 10^{-3} \le \beta \le 0.99$ and for
nucleon decay catalysis cross section smaller than $\sim 1 \, mb$.
\end{abstract}

\end{frontmatter}


%
%
%
%



 
\pretolerance=100
\normalsize

\section{Introduction}

Within the framework of Grand Unified Theories (GUT),
supermassive magnetic monopoles ($m \simeq 10^{17} \,$GeV) would 
have been produced in the early Universe
as intrinsically stable topological
defects when the symmetry of the unified
fundamental interactions was spontaneously broken 
\cite{preskill}.
At our epoch they should be searched for in the cosmic radiation as
remnants of primordial phase transition(s).
The velocity range in which GUT monopoles should be sought spreads over
several decades \cite{groom}.
If sufficiently heavy ($m \geq 10^{17} \,$GeV), GUT monopoles
will be gravitationally bound to the galaxy with a velocity distribution
peaked at $\beta \simeq 10^{-3}$. Lighter monopoles ($m \leq 10^{15}\,$GeV)
would be accelerated in one or more regions of coherent galactic magnetic 
field up to velocities of $\beta \geq 10^{-2}$ while other acceleration mechanisms 
(e.g. in the neighborhood of a neutron star) could bring them to 
relativistic velocities.

MACRO was a multipurpose underground detector (located in the Hall B of the 
Laboratori Nazionali del Gran Sasso, Italy) optimized for the search for GUT 
monopoles with velocity $ \beta = v/c \geq 4\cdot 10^{-5} $ and with a 
sensitivity 
below the Parker bound (i.e. $10^{-15} \,$cm$^{-2}$s$^{-1}$sr$^{-1}$ 
\cite{preskill,groom}).
The apparatus was arranged in a modular structure with overall dimensions of 
$76.5 \times 12 \times 9.3 \,$m$^3$ and was made up by three subdetectors:
liquid scintillation counters, limited streamer tubes and nuclear track 
detectors (CR39 and Lexan) \cite{primosm}.

In this work we describe an analysis technique which uses all of the 
MACRO subdetectors simultaneously in the  search
for fast moving magnetic monopoles (i.e. $\beta \ge 5 \times 10^{-3}$) 
in the cosmic radiation. 
The results on a first data sample is also reported.
The analysis procedure uses the data coming from the
scintillator and the streamer tube subdetectors to identify candidate events.
This is done by reconstructing the energy release (using scintillators' 
Energy Reconstruction Processor --ERP-- as well as streamer tubes'
Charge and Time Processor, QTP) and the particle's trajectory
(using the streamer tubes' digital hit information).
The selected tracks are
then searched for in the track-etch layers as a final tool for their
rejection or confirmation.

Since the three techniques are independent and complementary,
we obtain a good rejection power against the background
due to high energy muons and a high reliability of the possible candidates.
In the following sections we will describe the signatures of a fast monopole
in the MACRO detector and the analysis procedures.
Analysis efficiencies and acceptance evaluations will then be given together with
the results of a search performed on a first sample of more than two
years of data.
 
\section{Fast monopole signatures in MACRO}
\label{sec:signa}
A complete description of the MACRO detector, with particular emphasis on magnetic
monopole detection techniques, has been given elsewhere
\cite{primosm,stmono,phraserp}.
Here we will concentrate on the aspects which concern this particular analysis.

The energy loss suffered by a fast magnetic monopole in matter
is due to atomic excitation and ionization. Thus the average rate 
of energy loss follows immediately from an extension of the Bethe-Bloch formula
for a moving magnetic charge. This allows a straightforward and well-grounded
calculation of the detector response.
The peculiar characteristic of the fast magnetic monopoles is
their large ionizing power compared either to considerably slower monopoles or to 
minimum ionizing electrically charged particles.
The energy released in a MACRO streamer tube by a $\beta > 5 \cdot 10^{-3}$
monopole is at least one hundred times larger than that due to a minimum
ionizing particle (m.i.p.) \cite{gbation}.
Since the charge of the streamer pulse has been shown to have a logarithmic
dependence on the ionizing power \cite{gbation,laser}, it provides a good 
rejection  of the muon background.
The response of the scintillation counters to a magnetic monopole of a given
velocity is reported in \cite{ahlentarle,ficenec}. 
The expected signal for a monopole is about a 
factor 30 greater than the m.i.p. level for $\beta = 5 \cdot 10^{-3}$ and 
is even larger for greater velocities. 
In the case of the track-etch detectors,
the calculated value of Restricted Energy Loss (REL, defined as the fraction of 
the particle's energy loss contained in a 10$\,$nm diameter around its trajectory) 
for a magnetic monopole with $\beta=5 \cdot 10^{-3}$ is 
$\simeq$ 60 MeV cm$^2$g$^{-1}$, and
it increases rapidly at higher velocities \cite{derkaoui}.
This value is well above the detection threshold of the MACRO 
CR39 nuclear track detector, REL$_{min}$ = 26 MeV cm$^2$g$^{-1}$.
The Lexan has a much higher threshold compared to 
that of CR39, making it sensitive to relativistic monopoles only.

The particular signatures of a fast monopole in MACRO are thus:
$a$) a single track in the detector's spatial and temporal views
corresponding to a particle with a velocity greater than
$5~\cdot~10^{-3}~\,~c$, and
$b$) a large energy deposition along the particle trajectory.

The background for this search is almost entirely due to the flux of high 
energy single muons. At the depth of MACRO, about $5\%$
of these have energies greater than the threshold energy above which
the discrete energy loss processes of nuclear interactions, bremsstrahlung,
and pair production start to dominate, viz., $\sim 1 \, $TeV. These discrete 
mechanisms produce showering events in the apparatus.

This background is generally reduced by a combination of geometrical
and energy cuts imposed on each of the subdetectors.
An analysis based on all three subsystems in MACRO allows the use of
these cuts in a rather conservative way.
In particular, the energy cut remains safely within the linear response
of the system.
Moreover, any systematic errors are greatly reduced by the combination
of measurements from the three subsystems.

Energetic muons ($E_{\mu} \ge 1 \, TeV$) showering in the
apparatus are efficiently rejected by requiring a clean single track in
the streamer tube system.
However, this cut will also reject monopole events in
which electromagnetic or hadronic showers are produced.  In particular, a 
monopole with $\gamma \ge 10$ (i.e. $\beta \ge 0.99$)
can produce a $\delta$-ray in a single electron collision with enough energy to 
produce an electromagnetic shower, giving rise to secondary tracks.
Furthermore, processes such as hadronic interactions and $e^+ - e^-$ pair 
production are effective for $\gamma \ge 100$ \cite{ahlen78} and will produce 
secondary tracks.
Finally, another possible source of non-clean monopole-induced single tracks in the 
apparatus might be due to the nucleon decay catalyzed by monopoles 
following the Rubakov-Callan effect \cite{preskill}. 
Due to these effects the present analysis might lose efficiency in
identifying monopole events if the 
catalysis cross section $\sigma_{cat} \ge 1 \, mb$ or $\beta \ge 0.99$.

\section{Event reconstruction}

This search begins with event selection performed by both the ERP and the
Streamer tube Horizontal Monopole Trigger (SHMT).

Event reconstruction starts from the procedures used in the
search for slow monopoles with the streamer system \cite{stmono}.
Once event tracking has been performed, the energy loss observed in the
scintillators and the analog response of the streamer tubes are
used to achieve the necessary rejection power against the muon background.

\subsection{Event tracking in the Streamer Tube System}
\label{par:strea}
 
The details of the event tracking are fully described in
\cite{stmono}. The strategy adopted to eliminate the radioactivity
background is briefly summarized in the following.
The SHMT selects events with at least 7 horizontal planes
with a rough temporal alignment (a ``$\beta$ slice'').
For each of the triggered $\beta$-slices, wire hits are selected on the basis of
time measurements made by the QTPs, while strip hits are identified by
requiring a spatial match with the wire view.
If a track is found in each of the two spatial views, a more refined 
time track is searched for by using the QTP timing information from
the selected spatial hits.  
Furthermore we require a minimum of 7,6 and 7 points along the track in the wire,
strip and time views respectively.
Only events satisfying these requirements and consisting of a single track 
reconstructed in each of the wire, strip, and time views are
analyzed further.

\subsection{Measurement of the energy loss in scintillation counters}
\label{par:scint}

\begin{figure}[p]
  \begin{center}
 \mbox{\epsfig{figure=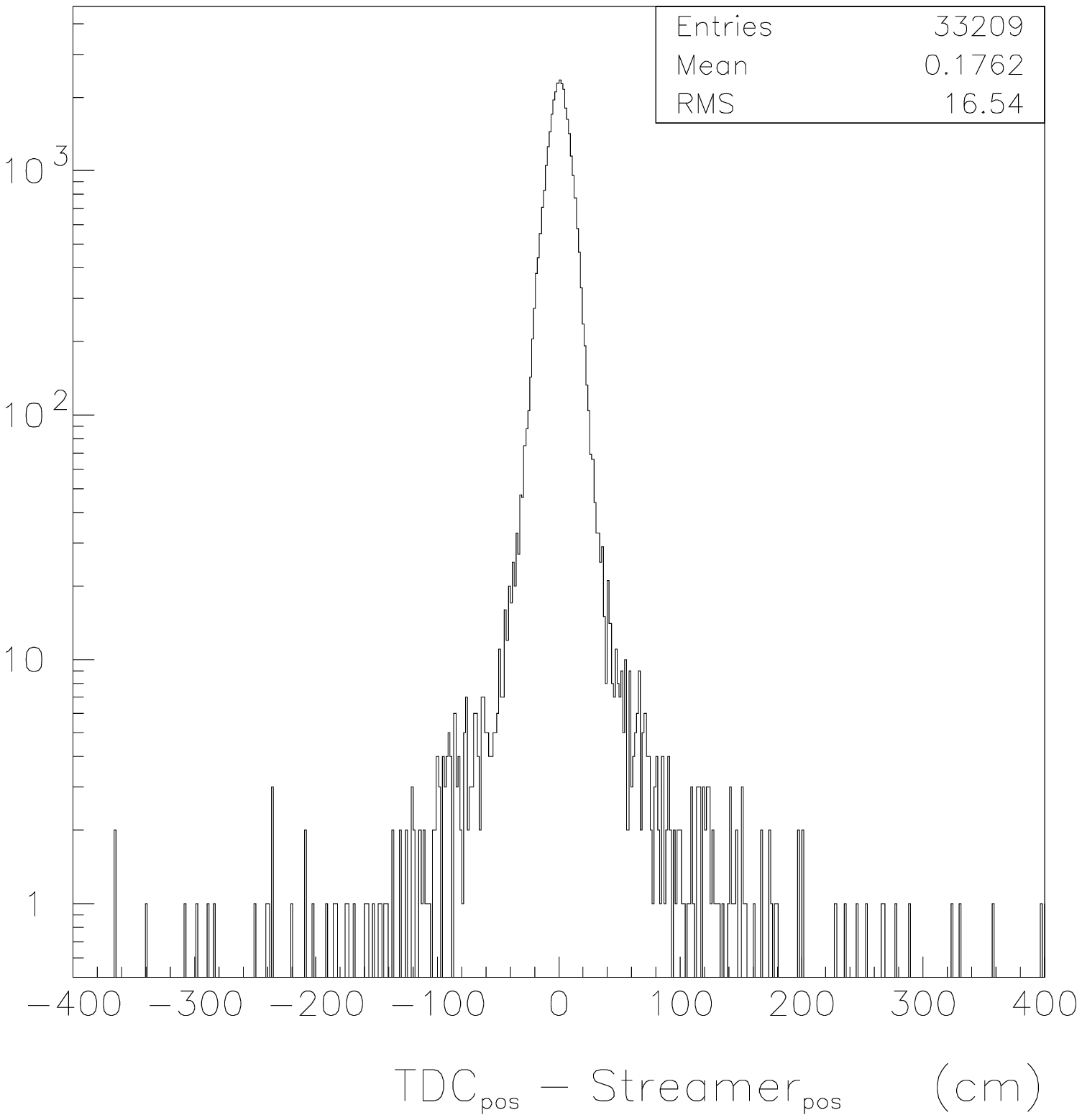,width=14cm,bbllx=31bp,bblly=200bp,bburx=550,bbury=630bp}}
  \vskip 1.cm
  \caption{\it \label{fig:dpos}
           Difference between the measurements of the position of the particle 
           crossing along the scintillator as given by the ERP TDC and by 
           the streamer tube tracking information.
           The time resolution of the scintillators comes out to be 
           $\sigma_t \simeq 750 \, ps$.
           In the analysis the hits for which this difference is greater than
           $100 \,$cm are not used.}
\end{center}
\end{figure}

\begin{figure}[p]
  \begin{center}
   \begin{tabular}{cc} 
\mbox{\epsfig{figure=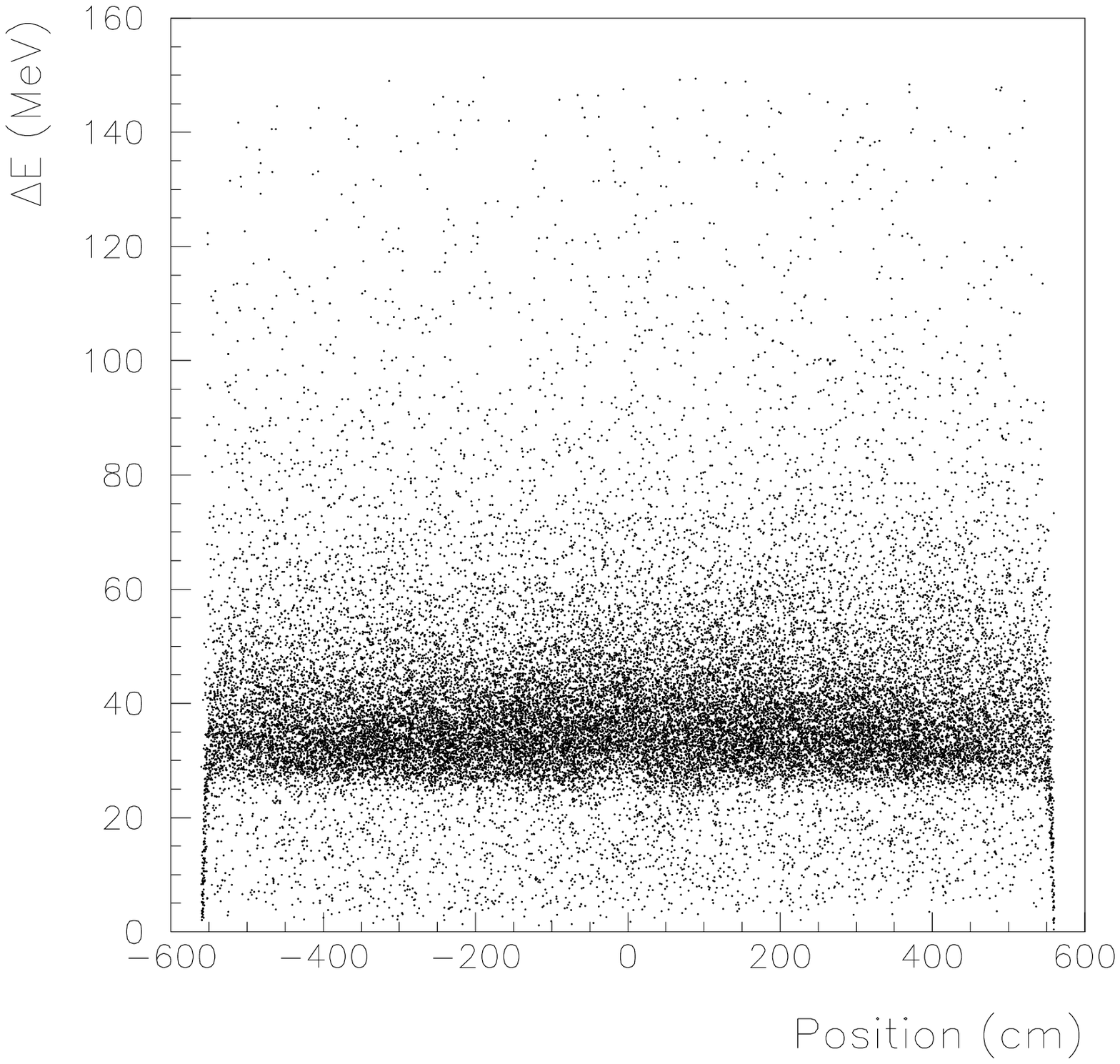,width=7cm,bbllx=31bp,bblly=200bp,bburx=550,bbury=630bp}}
   &
\mbox{\epsfig{figure=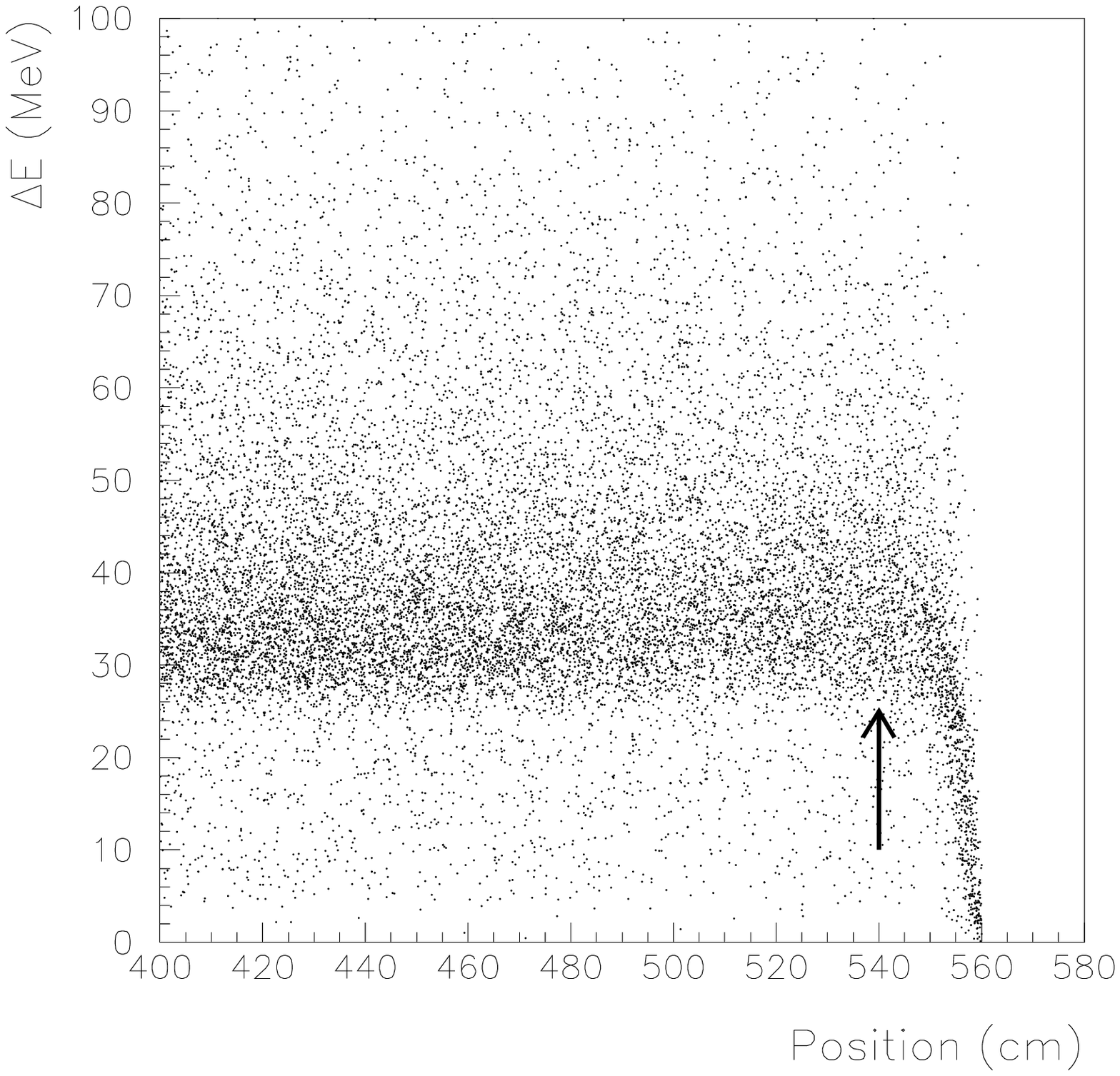,width=7cm,bbllx=31bp,bblly=200bp,bburx=550,bbury=630bp}}
   \end{tabular}
   \vskip 1.cm
 \caption{\it \label{fig:energy}
           Result of the energy reconstruction procedure in the sensitive
           volume. The reconstructed energy is independent on the position of the 
           particle crossing along the scintillator longitudinal axis. 
           On the right, the cut applied to exclude  corner clipping tracks and 
           border effects is also shown.}
\end{center}
\end{figure}

The ERP, a system measuring the energy deposition in each MACRO
single scintillator and time of flight across the apparatus, 
has been extensively described elsewhere
\cite{primosm,phraserp}.
In order to obtain $\Delta E$, the energy lost by a particle in a 
scintillation counter, the 
raw ADC values from each of the two tank ends are corrected for light attenuation in 
the scintillator and a normalized weighted mean of the two is taken.
The normalization constant for each scintillator is updated on a weekly basis
by fitting the Landau-like signal size distribution produced by a selected set of 
cosmic muon events.
Correcting for the light attenuation requires knowledge of the  longitudinal 
position of an event in a scintillator. This can be obtained
either from the streamer tube tracking system or from the difference of arrival 
times of the light at the two scintillator ends. In this
analysis, we use the former since it yields better energy resolution than the 
time-difference method.
In Fig.\ref{fig:dpos} the difference between the positions reconstructed  with
these two different methods is shown for a sample of cosmic muons.
The width of this distribution is primarily due to the time resolution of the
scintillators and it yields  $\sigma_t \simeq 750 \, ps$.
In Fig.\ref{fig:energy} the value of the reconstructed energy deposition of a sample 
of muons is shown as a function of the longitudinal position of the event 
in the scintillator.
As can be seen the correction for varying sensitivity as a function of this position
is performed correctly in almost all the sensitive volume.
Near the scintillator ends this procedure fails by underestimating the
energy lost by the particle. For this reason a fiducial volume is defined
by cutting the last $20 \,$cm close to each end.

A study of the response of the ERP to large signals has been fully described
in \cite{phraserp}. The response of the whole
chain from the photomultiplier tubes (PMT) to the recording analog-to-digital
converting (ADC) electronics was measured as a function of light level
by pulsing each individual counter with UV light coming from a nitrogen laser 
($\lambda \simeq 337 \,$nm) after passing through a precision light attenuator.
The system has been shown to have a linear response
up to a level corresponding to roughly 5 times the amount
of light produced in a pathlength of $\simeq 30 \,$cm by a minimum ionizing
muon. In this search the synergetic use of the digital (tracking) and analog
(pulse charge) information coming from the streamer tube system
allowed putting analysis cuts below the onset of ERP non-linearity.

Since in this search we are interested in monopoles with velocity down to
$\beta = 5 \cdot 10^{-3}$, the scintillator crossing time might exceed the width of 
the ERP integration gate. In this case only a fraction of the PMT pulse would be
integrated, thus underestimating the energy lost in the scintillator.
The ERP ADC integration gate is $200 \,$ns wide and the time spent by the
digital electronics to produce it is $70\,$ns. As a consequence the
PMT signal to be integrated is delayed by $75\,$ns.
Moreover the decay time of the slow component of the scintillation light,  the
electronic chain bandwidth, and the geometry of the detector actually set a
minimum pulse duration of about $60\,$ns, which can be estimated by observing
the muon pulses.
This implies that saturation effects might start for pulses produced by
particles which traverse the scintillator in more than $135 \, ns$.
However in $135 \, ns$, a $\beta = 5 \cdot 10^{-3}$ monopole travels about
$20 \, cm$ and loses at least $\simeq 465 \, MeV$ \cite{ahlentarle}.
Since the analysis cut will be set at a much lower level,
the width of the integration gate does not affect the detection efficiency.

The selection criteria adopted in this analysis for the ERP hits are then
summarized as follows:
\begin{itemize}
\item The values given by the ERP TDC's at the two different system thresholds
({\it high} and {\it low}) cannot differ, at each scintillator end, by more than 
$3\,$ns. This is essentially a check that the system is working properly.
\item The measurements of the longitudinal position of a hit as
performed by the streamer system tracking and by the ERP TDC's, cannot differ
by more than $100 \,$cm (see Fig.\ref{fig:dpos}).
\item A fiducial volume is defined to guarantee a proper energy reconstruction.
This is done by excluding $20 \,$cm close to the scintillator ends 
(see Fig.\ref{fig:energy}).
\item The pathlength of the particle in the scintillator volume,
as measured by using the streamer system track parameters, must be greater than
$10 \,$cm.
\end{itemize}
 
For all the events in which at least two different scintillators
are selected, the time of flight information is also available.

\subsection{Measurement of the streamer pulse charge}
 
\subsubsection{Study of the analog response of the streamer tubes}
\label{par:qtpsel}
Experimental studies of the dependence of the streamer charge on the
particle's trajectory and ionizing power are reported in
\cite{gbation,laser,cern}.
In this section we show the results of a study of the
MACRO streamer tubes charge, as measured by the QTP system,
using cosmic ray muons. This is done to better understand
the detector characteristics as well as  the dependence of the charge on the
geometry of the ionizing track, the gas mixture parameters, etc.
The understanding of these parameters allowed us to set up a
procedure for distinguishing a m.i.p. from a highly
ionizing particles such as a fast GUT monopole.
 
For each selected event, QTP hits along the track are
identified in order to measure the streamer pulse charge.
They must:
\begin{itemize}
  \item spatially match with the wire and strips hits;
  \item be in the time view within $\pm 2 \, \mu s$ from the temporal
        track;
  \item not be preceded in the previous $5 \, \mu s$ by another
        hit in the same QTP channel (this is required in order to both
        reject afterpulses and avoid systematic errors in the charge
        measurements due to the ADC's recovery time).
\end{itemize}
In the analysis we considered only wire hits with cluster size CLS=1 or 2
in a given QTP channel. In the following we will refer to the streamer
pulse charge as measured by the QTP receiving the signals from the horizontal
tubes, once the above mentioned selection criteria are applied.
 
\begin{figure}[p]
  \begin{center}
 
\mbox{\epsfig{figure=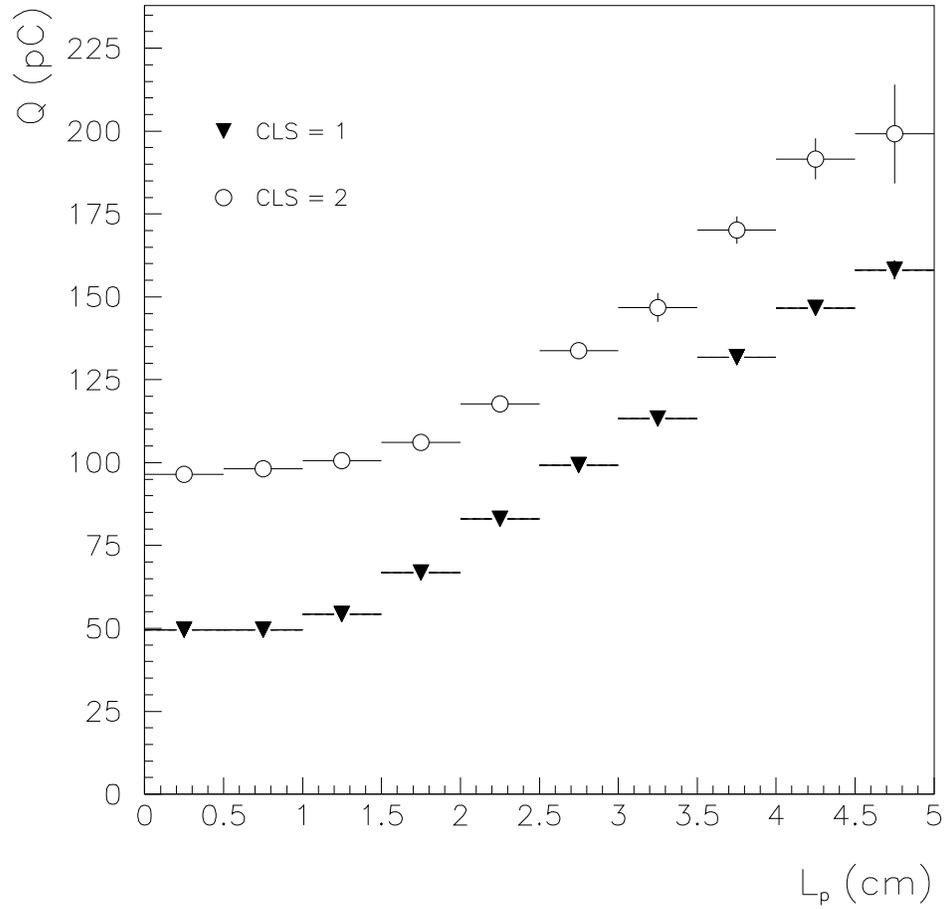,width=14.cm,bbllx=31bp,bblly=200bp,bburx=550,bbury=630bp}}
   \vskip 1. cm
   \caption{\it \label{fig:q_lp}
              Charge produced by cosmic muons in the MACRO horizontal
              streamer tubes as a function of $L_p$.
              The two plots refer to CLS=1 and CLS=2 data.}
  \end{center}
\end{figure}

For slanted particle trajectories with respect to the direction of the anode
wire, the probability to produce more than one streamer in the gas volume
increases.
As a consequence, after a region (which we refer to as the dead zone) in which
the multi-streamer production is strongly reduced by space charge effects due
to the first avalanche, the pulse charge linearly increases for
increasing $L_p$, the length of the projection of the particle track
(in the sensitive volume) along the direction of the wire.
This dependence has been studied in reference \cite{cern} using a muon beam. Here we
report also the results of a study on the MACRO streamer tubes performed with
cosmic muons crossing the detector.
In particular, for each $0.5 \,$cm interval in $L_p$, the charge
distributions produced by cosmic muons have been fit.
Gaussian fits have been separately applied for CLS=1 and CLS=2 hits.
The mean value of the fits are given as a function of  $L_p$ in
Fig.\ref{fig:q_lp}.
These behaviors are in good agreement with the aforementioned results obtained
in \cite{cern}. The dependence of the charge on $L_p$ comes out
to be:
\begin{eqnarray}
 \label{eq:normaq}
 Q = Q_0 &                        & ~~~~ \, {\rm if} ~~~ L_p \leq L_d \\
 \nonumber
 Q = Q_0 &~+~ S \cdot (L_p - L_d) & ~~~~ \, {\rm if} ~~~ L_p \geq L_d 
\end{eqnarray}
where $Q_0$, $S$ and $L_d$ are the values of the charge produced by
vertical muons, the slope of the linear part of the plot, and the width of the
dead zone, respectively.
Identical behaviors are obtained with vertical streamer tubes.
Only the results for $L_p \leq 5\,$cm are reported.
This is due to the angular distribution
of the cosmic muons that limits the statistics for $L_p \geq 5\,$cm.
It must be noticed that, given the trigger requirements of the SHMT,
the restriction to the analysis of events with $L_p \leq 5\,$cm
results only in a $6\%$ reduction of the geometric acceptance.
 
\begin{figure}[hp]
  \begin{center}
 
\mbox{\epsfig{figure=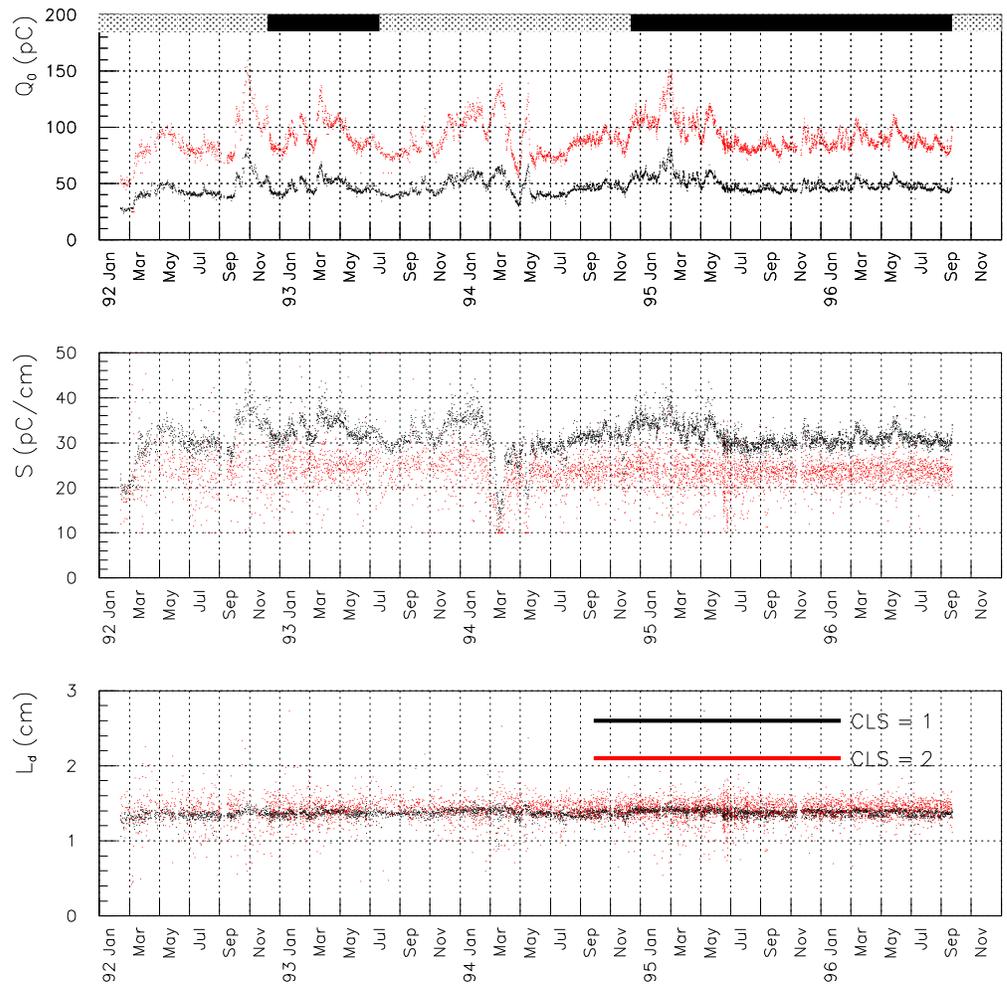,width=14cm,bbllx=31bp,bblly=150bp,bburx=550,bbury=800bp}}
   \vskip 0.5 cm
   \caption{\it \label{fig:calibrast}
             $Q_0$, $S$ and $L_d$ as a function of time,
             for CLS=1 (black points) and CLS=2 (grey points) hits.
             The black strips on the top indicate the periods used
             for this search.}
  \end{center}
\end{figure}

The procedure used to obtain the plots shown in Fig.\ref{fig:q_lp} has been
applied to the whole data sample.
For calibration purposes the values of $Q_0$, $S$ and $L_d$ have been measured
on a run-by-run basis. This allowed us to study and monitor the MACRO streamer
tube  response, its dependence on the gas mixture parameters and the
correlation between the CLS=1 and CLS=2 values.
The behaviors of $Q_0$, $S$ and $L_d$, for CLS=1 and CLS=2, are reported
in Fig.\ref{fig:calibrast} as a function of the solar time.
The fluctuations in the value of the charge produced by vertical muons
are due to small variations of the parameters of the gas mixture.
In March 1994, during hardware work for detector upgrade,
there was an abnormal streamer charge response (see Fig.\ref{fig:calibrast}),
due to a strong air contamination of the gas mixture.
This caused the tubes to operate in Geiger mode.
 
\begin{figure}[hp]
  \begin{center}
 
\mbox{\epsfig{figure=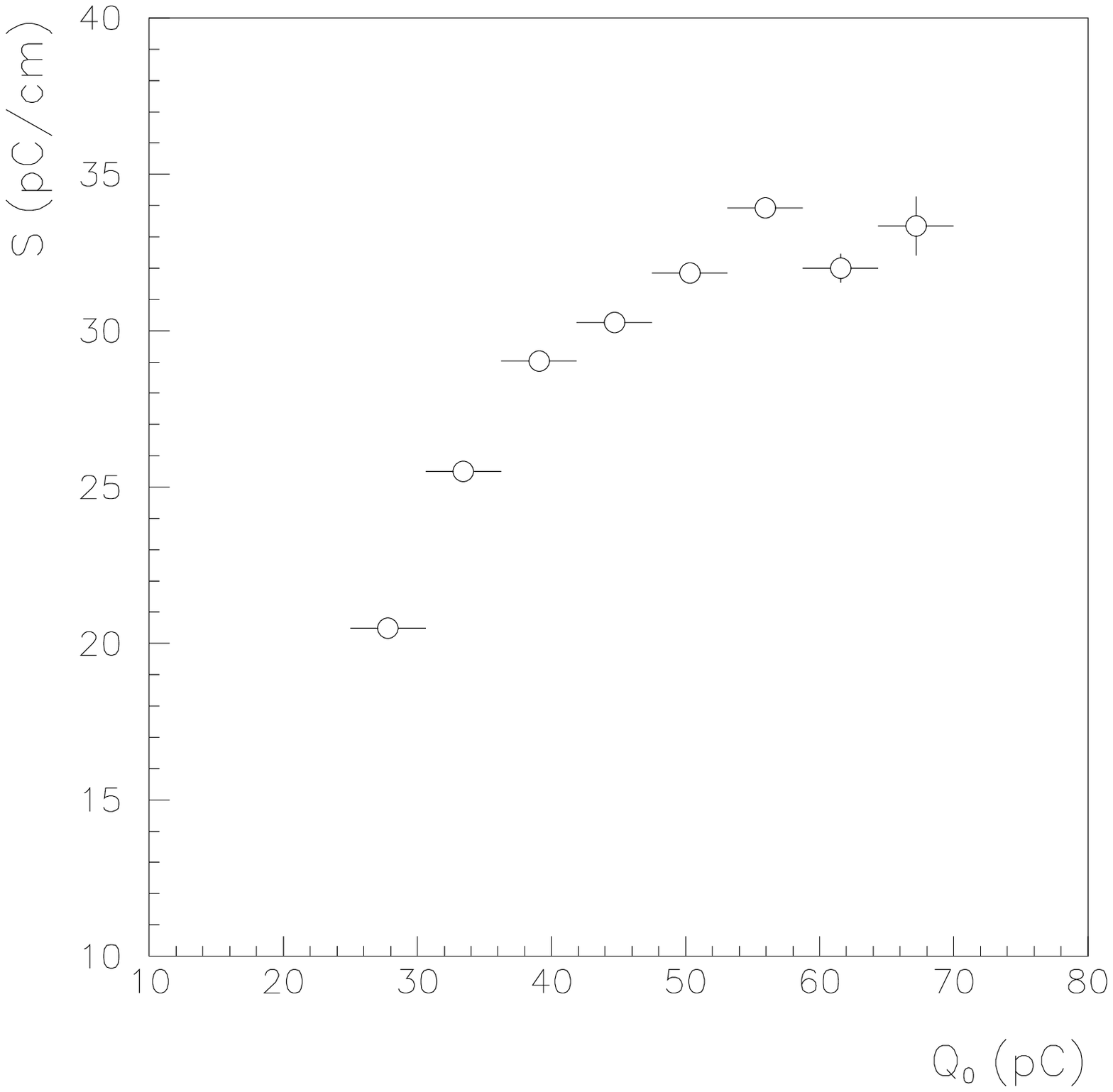,width=12cm,bbllx=31bp,bblly=200bp,bburx=550,bbury=630bp}}
   \vskip 0.5cm
 
\mbox{\epsfig{figure=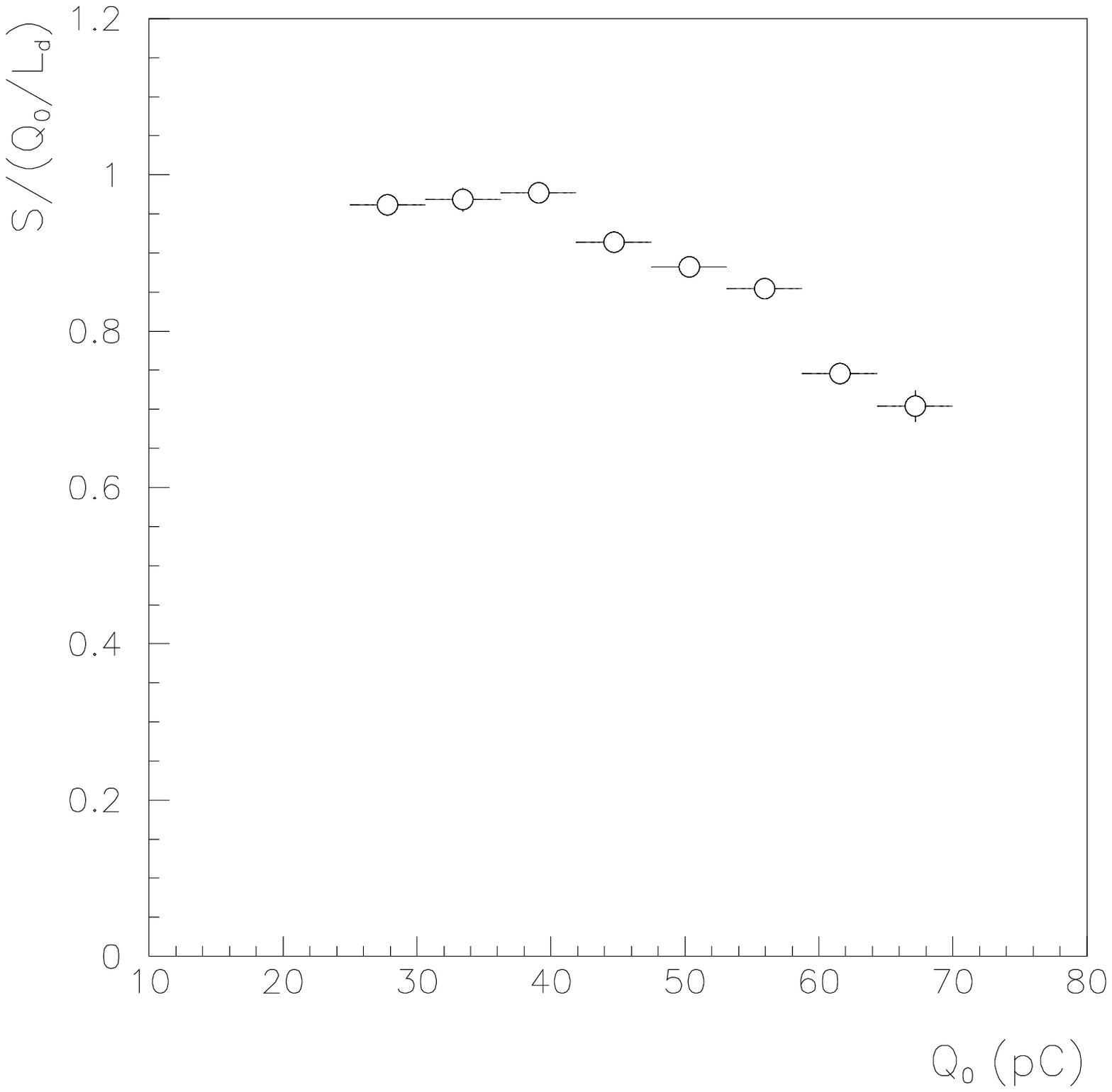,width=12cm,bbllx=31bp,bblly=200bp,bburx=550,bbury=630bp}}
   \vskip 1.cm
    \caption{\it \label{fig:slo_q}
            Dependence of $S$ on $Q_0$ for CLS=1 hits in the horizontal
            tubes. This behavior is also compared with
            the ratio $Q_0/L_d$.}
  \end{center}
\end{figure}

Small variations in the actual values of $Q_0$, $S$ and $L_d$
are very useful in order to study their mutual dependencies and to
compare them with the results obtained in the literature \cite{cern}.
One example is the dependence of $S$ on $Q_0$,
which is shown in Fig.\ref{fig:slo_q} for CLS=1 hits.
Since the increase of the pulse charge for larger values of $L_p$ is due
to the production of more than one streamer avalanche along the particle
trajectory, one should expect $S \sim Q_0/L_d$.
However, in agreement with what is reported in \cite{cern}, this
approximation fails for large values of $Q_0$, when it is not easy to
take into account space charge effects in a simple phenomenological model.
The important feature of the first plot of Fig.\ref{fig:slo_q} is that
for a highly ionizing particle (i.e. large $Q_0$) the value of $S$ has
to be larger than the corresponding value for a m.i.p. in the same detector
configuration.
This is particularly important for this analysis
since it allows a better discrimination between fast monopoles and
cosmic muons.

\subsubsection{Measurement of the average charge along the track}
\label{par:gamma}
The results of the study of the MACRO streamer tube response
and the aforementioned calibration procedure allowed us to
correct the pulse charge measurements for both geometrical effects
and gas mixture variations.
 
The charge measured by the QTP system for the selected hits has been
normalized by dividing it by the value of the average charge produced
by a muon in the same operational and geometrical conditions.
The muon charge is calculated by using Eq.\ref{eq:normaq} with the value of
$L_p$  for the event under consideration and the values of $Q_0$, $S$ and $L_d$
from the calibrations for a given CLS and run.
As can be seen in Fig.\ref{fig:calibrast}, the values of $Q_0$,
$S$ and $L_d$ for CLS=2 have large spreads around their mean values. 
This is a purely statistical effect in the fit to the various charge distributions.
However the ratios of the CLS=2 constants to the CLS=1 ones are very stable in
the whole data set. As a consequence, we avoided small
effects due to the aforementioned spreads by using the values obtained
for CLS=1 multiplied by the measured scaling factor.

In order to measure the streamer tube response to the passage of a
particle, we defined a variable $\Gamma \equiv Q/Q_{\mu}$, where
$Q$ is the charge actually measured for a given event and $Q_{\mu}$ is that
produced by a muon and calculated as described above.
As was reported in \cite{gbation,laser} $\Gamma$ has a logarithmic
dependence on the particle ionizing power. In particular, for a
monopole with $\beta \ge 5 \cdot 10^{-3}$, $\Gamma \ge 4$ is expected.
The value of $\Gamma$ has been measured in each selected event by using the QTP
information coming from the horizontal streamer tubes for CLS=1 and CLS=2 hits
\footnote{For CLS $>$2 the poor statistics (due to the muon angular
distribution) does not allow setting up a calibration procedure on a run
by run basis.}.
In this way $n$ different measurements for CLS=1 ($\Gamma^i_{CLS=1}$)
and $l$ different measurements for CLS=2 ($\Gamma^i_{CLS=2}$) have been
obtained along the particle trajectory.
The value of $\Gamma$ associated with the event is then obtained by
averaging these $n+l$ measurements, weighted by the inverse of the variances
of the charge distributions,
$\sigma^2_{CLS=1}$ and $\sigma^2_{CLS=2}$, as obtained from the data
sample under consideration.
In order to reduce the fluctuations of the value of
$\Gamma$, we do not include in the average the hit with the largest
value of $\Gamma_{CLS=s} /\sigma^2_{CLS=s}$ ($s$=1 or 2).
Therefore
\begin{equation}
 {\small
 \label{eq:mediagamma}
 \Gamma = \frac{ \sum_{i=1}^n \Gamma^i_{CLS=1} /\sigma^2_{CLS=1}  +
 \sum_{i=1}^l \Gamma^i_{CLS=2} /\sigma^2_{CLS=2} -
 (\Gamma_{CLS=s} /\sigma^2_{CLS=s})_{max} }
 { \sum_{i=1}^n 1/\sigma^2_{CLS=1}  + \sum_{i=1}^l 1/\sigma^2_{CLS=2} -
 1/\sigma^2_{CLS=s}  }
 }
\end{equation}
which is calculated on $ n_{\Gamma} = n + l - 1$ different points.
Given the considerations concerning the efficiency and
the acceptance of the analysis (see next sections), we required an event
to have at least $n_{\Gamma}=6$.

\begin{figure}[hp]
  \begin{center}
 
\mbox{\epsfig{figure=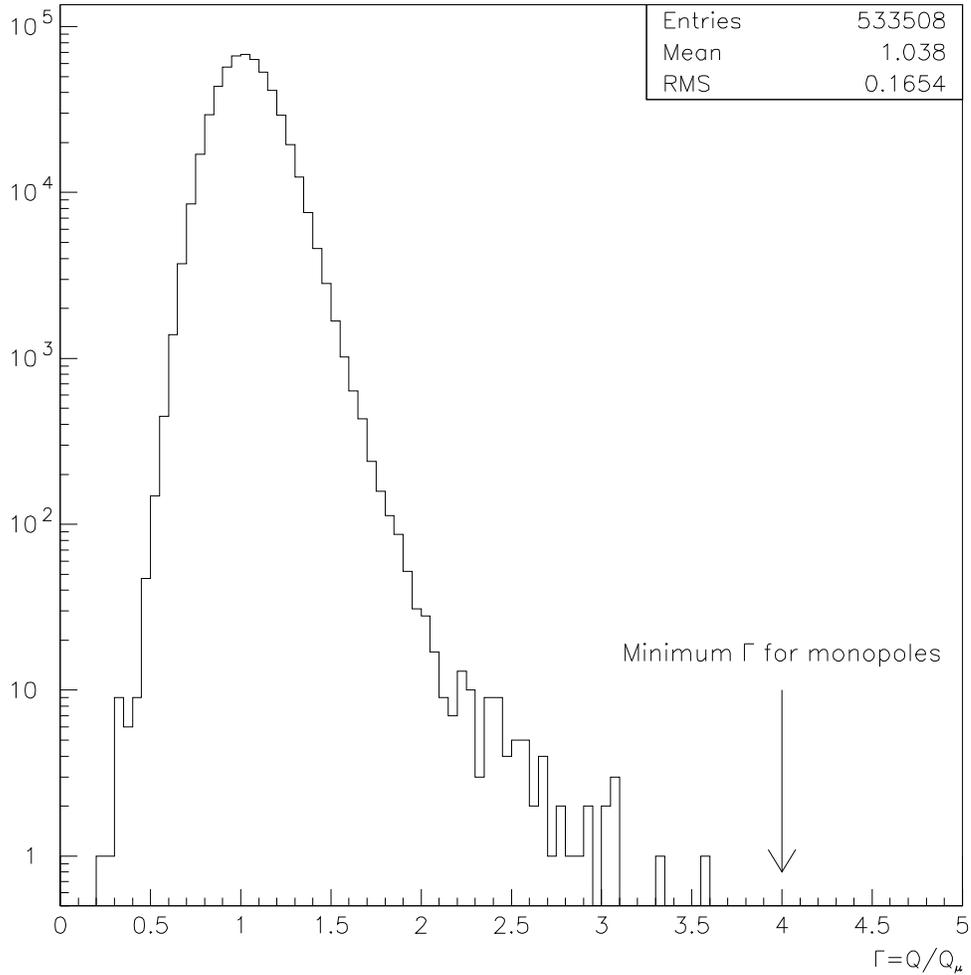,width=14cm,bbllx=31bp,bblly=200bp,bburx=550,bbury=630bp}}
    \vskip 1.cm
    \caption{\it \label{fig:gamma}
          Distribution of the values of $\Gamma$ for a data sample.
          A monopole with $\beta \ge 5 \cdot 10^{-3}$ would have
          $\Gamma \ge 4$.
          No cuts on the energy deposition in the scintillators has been
          applied at this stage.}
  \end{center}
\end{figure}

The distribution of $\Gamma$ for a data sample is shown in Fig.\ref{fig:gamma}.
As expected, the shape is almost Gaussian with a tail towards the largest
values that is due to both fluctuations and $\delta$ ray production by high 
energy muons.
As anticipated, a magnetic monopole with $\beta \ge 5 \cdot 10^{-3}$
would have $\Gamma \ge 4$. However, detailed studies of  the streamer
formation process and of the effects of the read-out electronics on the pulse shape 
show that the actual value of the cut must be a suitable
function $\Gamma_{cut}(L_p)$ as discussed in the following.

\section{Analysis flow}

The analysis method can be summarized as a sequence of three different phases:
run selection, event selection and reconstruction, and candidate selection.

In the first phase all runs longer than 0.5 hours are selected
(the average time length of a MACRO run is 6.5 hours). Evident electronic 
problems are ruled out by discarding runs with abnormal trigger rates.
This selection results in a total live time of more than $80 \,\%$ of the solar 
time.

Event selection requires that both the SHMT and the ERP trigger have fired.
Then single tracks must be identified in the spatial and temporal views
with the requirements listed in Sec.\ref{par:strea}.
Given a space track, at least one scintillator must satisfy
the selection criteria discussed in Sec.\ref{par:scint}.
For each of the selected ERP hits, the energy reconstruction
procedure is applied. If at least two different scintillators fired, 
a precise time of flight measurement (at the level of less than 1.ns)
is also possible.
The QTP hits for the streamer pulse charge measurements are then selected
(see Sec.\ref{par:qtpsel}). This allows the measurement of 
$\Gamma^i_{CLS=s}$ for all CLS=1 and CLS=2 wire hits in the horizontal 
streamer tubes, by using the streamer charge calibration constants 
previously produced for each run.
The calculation of the $\Gamma$ of the event is then made by making an average 
on at least $n_{\Gamma} = 6$ QTP hits (see Sec.\ref{par:gamma}).
The average amount of events reconstructed following these requirements is
$\sim 3 \cdot 10^6 \,$evt/yr.

Monopole candidate events are then selected. The energy 
reconstructed in each selected scintillator counter must be 
$\Delta E ~ \geq ~ (\Delta E )_{min} = 150 \,$MeV (see Sec.\ref{par:erpcut}). 
On the $\sim 10^3 \,$evt/yr which are left a cut on $\Gamma$ is imposed: 
$\Gamma ~ \geq ~ \Gamma_{cut} (L_p)$ (see Sec.\ref{par:gamcut}).
Finally the analysis of the possible survived candidates ($\sim 3 \,$evt/yr) 
is made by the scanning of the track-etch sheets identified by the streamer track.

\section{Efficiency evaluations}
\label{par:effi}
 
As can be envisaged from the previous sections, muon events
pass the first two analysis steps
but not the last, which is based on energy loss measurements in the 
scintillator, streamer and track-etch subdetectors. 
This allowed the use of the cosmic muons as an important tool for an evaluation
of the analysis efficiency.
The data from \cite{laser,cern} and some characteristics of the
streamer tube read-out card have been used to compute the
efficiency of the cut on the analog streamer response.

\subsection{Streamer tube system: trigger and event reconstruction efficiencies}
The SHMT sensitivity to muons allows a simple and direct measure of the
overall reconstruction efficiency $\varepsilon_{st}$ of the analysis steps based on 
the streamer information.
It has been estimated according to the procedure fully discussed in \cite{stmono}.
Basically $\varepsilon_{st}$ is expressed as
the product of the efficiency of the reconstruction algorithm
$\varepsilon_{rec}$ times the efficiency $\varepsilon_{trg}$ due to the
trigger and all the electronics used to produce the streamer data:
$ \varepsilon_{st} =  \varepsilon_{rec}  \cdot \varepsilon_{trg}$.
Actually the two contributions are not independent, leading to a conservative
estimate of the overall efficiency.
The reconstruction efficiency $\varepsilon_{rec}$ for particle
with $\beta \simeq 1$ is measured by the ratio between the rate of muons 
reconstructed by the analysis and the expected value. 
Then a check is performed of the independence of $\varepsilon_{rec}$ on 
the velocity of the particle. This is done by allowing the   
natural radioactivity hits to simulate time alignements corresponding to
slow particles.
Electronic failures that can be responsible for trigger inefficiency, i.e.
$\varepsilon_{trg}$, 
are monitored by computing the ratio between the measured trigger rate and the 
expected one on the basis of the knowledge of the trigger circuitry and of the 
radioactivity background rate.

\subsection{The ERP efficiency}
The scintillator counters considered in a given event are identified by
using the tracking information provided by the streamer tubes.
The single scintillator efficiency $\varepsilon_{scin}$, with respect to the event
reconstruction procedure with the streamer tubes, is measured for each run and 
each counter by checking if the ERP trigger associated with the tanks
crossed by the reconstructed streamer track actually fired. 
This single scintillator efficiency, $\varepsilon_{scin}$, is very close 
to one except for a few counters with evident hardware problems.
The ERP efficiency $\varepsilon_{ERP}$,  has then been evaluated
for each run by measuring the fraction of scintillators with
$\varepsilon_{scin} \ge 95 \% $.
This procedure is very sensitive to the overall ERP efficiency and it 
automatically takes into account all the inefficiencies of the tracking algorithm.

\subsection{The ERP energy cut}
\label{par:erpcut}
As calculated in \cite{ahlentarle}, a magnetic monopole with
$\beta \geq 5 \cdot 10^{-3}$ produces a light
yield in a scintillator which is more than 30 times larger than that due to the 
passage of a m.i.p.
If we take into account that this estimate has a factor of two
uncertainty \cite{ahlentarle}, the minimum light yield produced by a monopole
with the shortest accepted pathlength in the fiducial scintillator volume
($10 \,$cm) is
\begin{equation}
 \Delta E _{\cal M} \ge 15 \times 1.8 \, \frac{MeV}{g/cm^2} \cdot
  \rho_{sc} \cdot 10 \, cm \ge 232 \, MeV
\end{equation}
where $\rho_{sc} \simeq 0.86 \, g/cm^3$ is the mass density of the
scintillating liquid.
For this value of $\Delta E$, the non linearity effects (due to the
PMT and/or ADC saturation) are below the $20\%$ level and the energy 
resolution is about $5\%$ \cite{phraserp}. Therefore a cut which imposes
$\Delta E \geq 150 \,$MeV has an efficiency greater than $99\%$.

\subsection{The streamer charge cut}
\label{par:gamcut}
As shown before, a monopole with $\beta \simeq 5 \cdot 10^{-3}$ would produce
a track in the streamer system with $\Gamma \simeq 4$.
The streamer charge would be even larger for faster monopoles.
The analysis requires $ \Gamma \geq \Gamma_{cut}(L_p) $. The suitable value of
$\Gamma_{cut}(L_p)$ is chosen to ensure an efficiency of  at least
$90 \%$ for recognizing a monopole.
To evaluate $\Gamma_{cut}(L_p)$, a complete and detailed simulation has
been performed of the full analysis chain starting from the production of the
electronic signal.
To ensure that the input waveforms for the simulations were correct, we collected  
$80\,$MHz WFD data, at different $L_p$
values, from a streamer tube operating in a muon beam \cite{cern} under conditions
very similar to those in MACRO.
 
\begin{figure}[p]
  \begin{center}
 
\mbox{\epsfig{figure=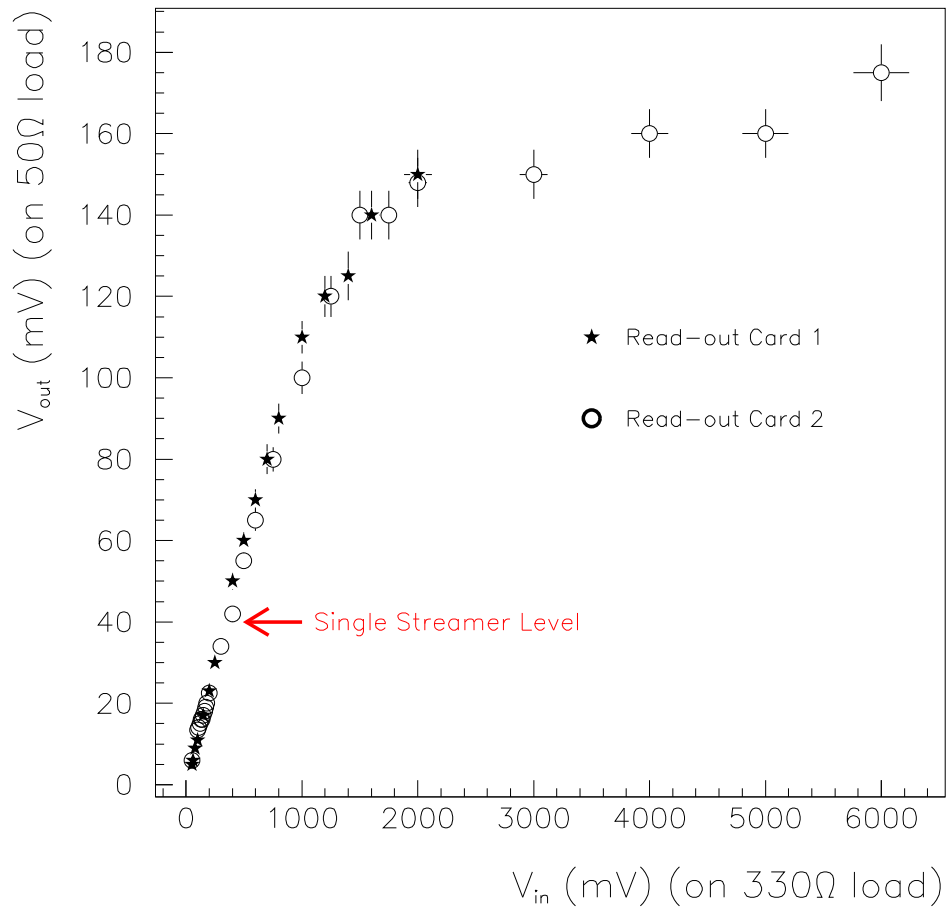,width=14cm,bbllx=31bp,bblly=200bp,bburx=550,bbury=630bp}}
   \vskip 0.5 cm
    \caption{\it \label{fig:diodo}
           The effect of the protection diodes of the read-out card is shown.
           The output amplitude saturates at a level corresponding to three
           times the pulse produced by a muon.}
  \end{center}
\end{figure}

Since $a)$ a monopole with $\beta \simeq 5 \cdot 10^{-3}$ is expected
to produce streamer pulses with a charge four times higher than that produced by a
muon, $b)$ the dead region length $L_d$ is essentially independent on $Q_0$,
and $c)$ the slope $S$ increases with $Q_0$, the monopole charge parameters
are conservatively set at:
\begin{eqnarray}
  \label{eq:qmono}
    \nonumber
    Q_0^{\cal M} &=& 4~ Q_0^{\mu}  \\
    L_d^{\cal M} &=&  L_d^{\mu}   \\
    \nonumber
    S^{\cal M}   &=&  S^{\mu}
\end{eqnarray}
In order to simulate the waveform of the signal produced by a
monopole, the muon streamer response was multiplied by the ratio of the charge 
produced by a monopole and that
by a muon as computed with Eq.\ref{eq:normaq} and Eq.\ref{eq:qmono}.
In the simulation, the effect of the protection diodes of the read-out cards is
also considered. They actually introduce an amplitude  saturation of wire
signal outputs corresponding to more than three times the signal produced by a
m.i.p. (see Fig.\ref{fig:diodo}). For each  $L_p$ value both muon and monopole
waveforms have been conservatively saturated at a  level corresponding to
two times the amplitude of the  single streamer pulse.
 
The values of $\Gamma$ have been determined, separately for  muon and monopole
events, by using the charges of the simulated pulses at different $L_p$.
For this purpose a  detailed Monte Carlo simulation of the MACRO detector
provided the number of hits with CLS=1 and CLS=2 for muons and monopoles, taking 
into account the different angular distributions and the
different effects of the  trigger requirements for monopoles and muons.

\begin{figure}[p]
  \begin{center}
 
\mbox{\epsfig{figure=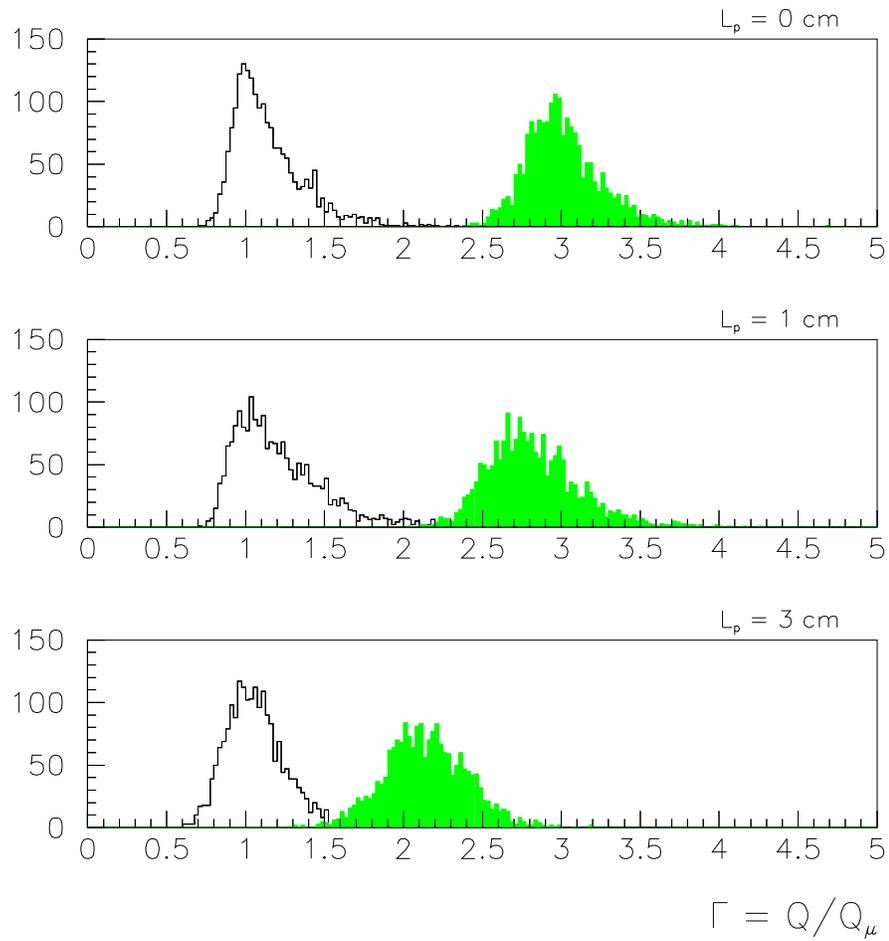,width=14cm,bbllx=31bp,bblly=200bp,bburx=550,bbury=630bp}}
   \vskip 1 cm
    \caption{\it \label{fig:mumono}
             $\Gamma$ values for simulated muons and $\beta = 5 \cdot 10^{-3}$
             monopoles (in grey) for three different $L_p$ values.
             The effect of the saturation of the pulse amplitude introduced
             by protection diodes (see text) is clearly visible.}
  \end{center}
\end{figure}

A result of this procedure for $\beta =5 \cdot 10^{-3}$ monopoles is shown
in Fig.\ref{fig:mumono}.  The $\Gamma$ distributions of muons and monopoles
are well separated for small $L_p$ while the effect of the
saturation due to diodes is clearly visible for larger $L_p$.
Starting from these data, a function $\Gamma_{cut}(L_p)$
was extracted for which at least $90 \%$ of monopoles
have $\Gamma \geq \Gamma_{cut}$.
For monopoles with $\beta > 5 \cdot 10^{-3}$, a larger value of $\Gamma$ is expected.
However, in the whole $\beta$ range covered by this analysis, we conservatively used
the $\Gamma_{cut}(L_p)$ estimated for $\beta =5 \cdot 10^{-3}$ as shown above.

\section{Analysis results on a first data sample}
\label{par:runse}
The analysis procedure described above has been performed on a first data sample
collected since December 1992 to September 1996, with the exclusion of the period 
June 1993-December 1994, when the efficiency of the streamer and scintillator 
subdetectors were affected by the works due to the upgrading of the apparatus 
with the construction of the upper part.
In particular, by considering the ERP and SHMT rates,
the ERP efficiency and the calibration of the streamer charge response we have 
selected three data taking periods with a total live time of more than 16,000 hours. 
The main characteristics of the considered data set, as well as  the values of the 
analysis efficiencies discussed above, are listed in Tab.\ref{tab:period}. 

\begin{figure}[hp]
  \begin{center}
    \mbox{\epsfig{figure=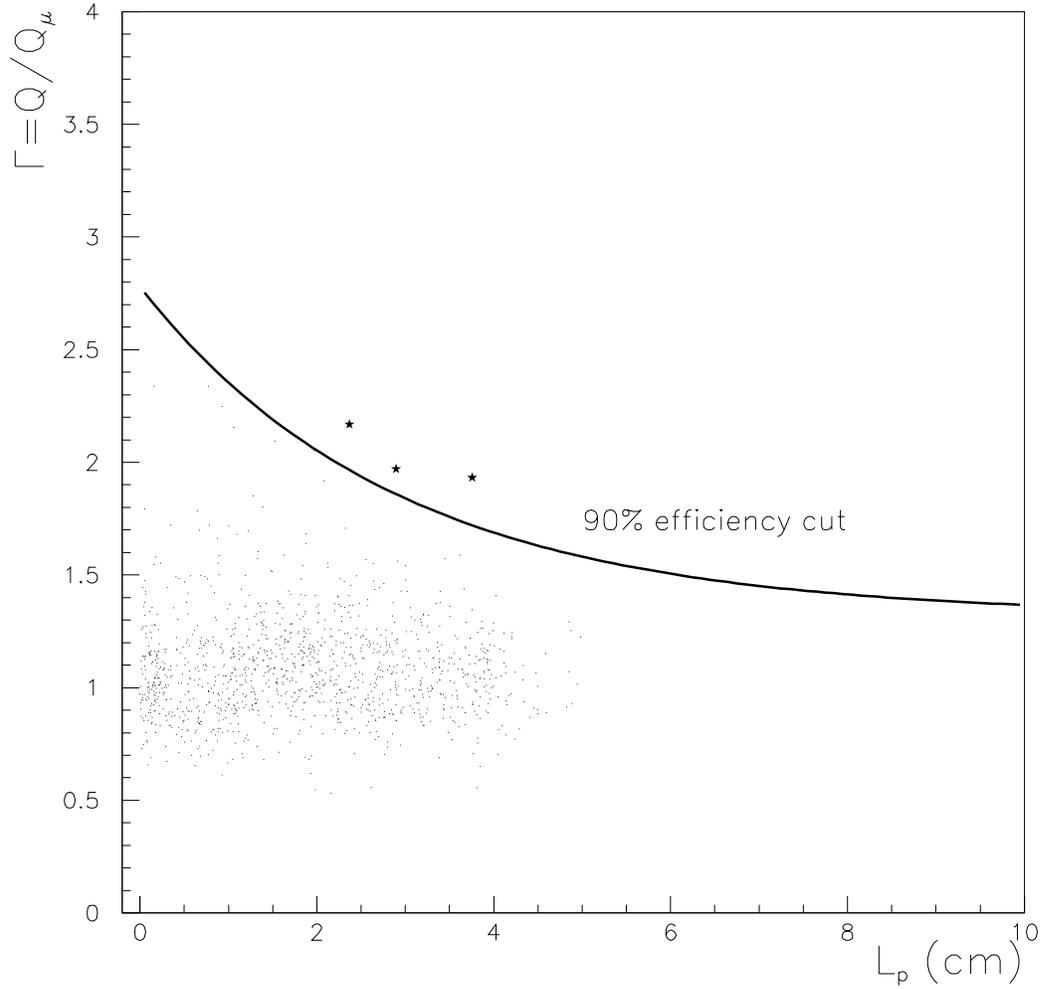,width=14.cm,bbllx=31bp,bblly=200bp,bburx=550,bbury=630bp}}
    \vskip 1.cm
    \caption{\it \label{fig:lpcut123}
              Value of $\Gamma$ as a function of $L_p$, for the
              surviving events after the $\Delta E \geq 150\,$MeV cut. 
              The stars indicate the candidates
              which passed the cut on $\Gamma$ (see text).}
  \end{center}
\end{figure}

\begin{figure}[hp]
  \begin{center}
   \mbox{\epsfig{figure=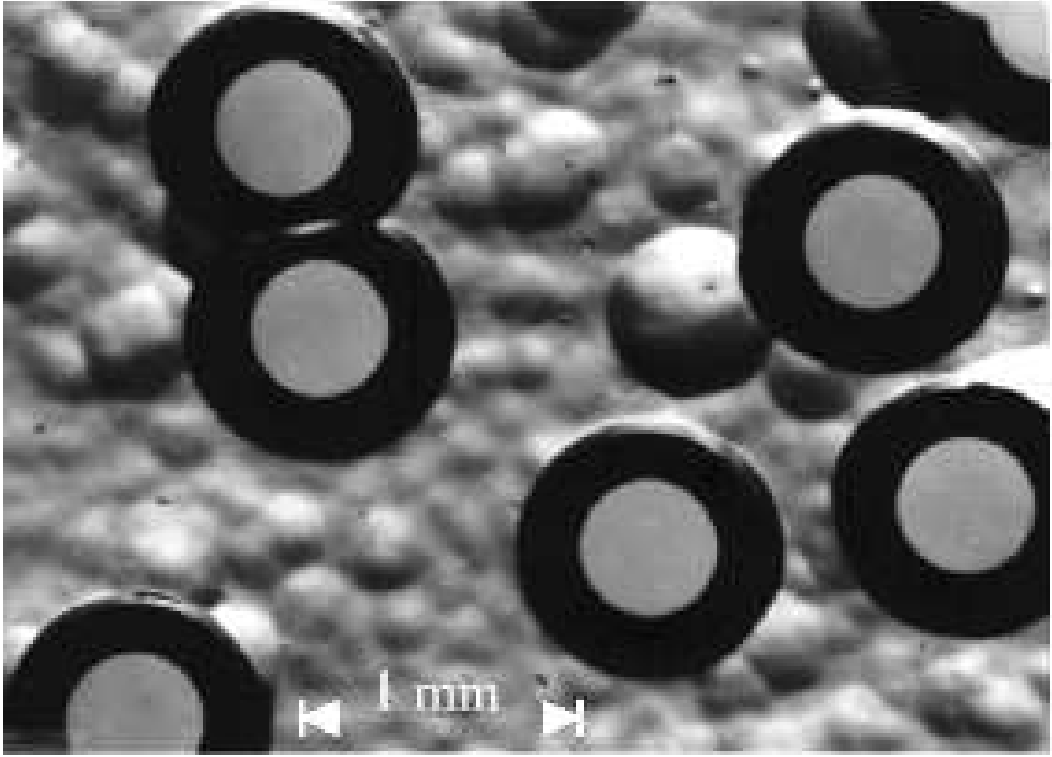,width=14cm}}
   \vskip 1. cm
    \caption{\it \label{fig:tracketch}
            Microphotographs of holes in 
            CR39 from  relativistic S$^{16+}$ ions after 120 h of etching in 
            NaOH 8N at 80$^{\circ}$ C. 
            The sulphur holes are clearly distinguishable from 
            the surface background.}
  \end{center}
\end{figure}

Once all the selected events were fully reconstructed, the
150 MeV energy cut mentioned above was applied. Then the streamer
charge on the surviving events was analyzed.
The result is shown in Fig.\ref{fig:lpcut123} where the 
dependence of $\Gamma$  on $L_p$ is reported together with the function 
$\Gamma_{cut}$ previuosly described.
\begin{table}
 \begin{center}
  \begin{tabular}{|c|c|c|c|c|}
  \hline
Period & Runs &  $\varepsilon_{ERP}$ & $\varepsilon_{st}$  &
$T_{live} ({\rm hours})$  \\
  \hline
  \hline
11 Dec 92 - 13 Jun 93 & 5520 - 6329 & 0.911 & 0.788 & 3456 \\
12 Dec 94 - 24 Jul 95 & 8907 - 10556& 0.915 & 0.894 & 3738 \\
24 Jul 95 - 23 Sep 96 &10557 - 12816& 0.915 & 0.946 & 8816 \\
  \hline
  \end{tabular}
 \end{center}
 \vskip .5cm
 \caption{\it \label{tab:period}
          Main characteristics of the analyzed data sample.
          The values of the analysis efficiencies (see Sec.\ref{par:effi})
          and the integrated live times are also reported.}
\end{table}
In the first period, three events are close to the cut, and, in the third period,  
three others survive the cut. For these six events the
track-etch wagons identified by using the streamer tracking system were extracted. 
The passage of a magnetic monopole would cause a structural damage. 
Chemical etching should result in the formation of collinear etch-pit 
cones of  equal size on both faces of each foil, or of a through-hole, 
after a prolonged etching. 
Fig.\ref{fig:tracketch} shows microphotographs of holes obtained by exposing a 
CR39 foil to relativistic sulphur ions (REL $\sim$ 300 MeV cm$^2$g$^{-1}$)
impinging normally to the detector surface. 
A candidate track must satisfy a three-fold coincidence of the position, 
incidence angles among the layers and should also give the same value of REL.
A detailed description of the analysis procedure is given in \cite{lp}.
For each extracted
wagon the top layer of CR39 was etched in NaOH 8N at 80$^{\circ}
\,$C for $\sim 120 \,$ hours. 
No track compatible with the crossing of a monopole was found.
Since the detector threshold is at more than 9 standard deviations below the expected
REL for a monopole with $\beta \simeq 2 \cdot 10^{-3}$, the
efficiency of the track-etch analysis can be considered $100 \%$ in this search.

\begin{figure}[hp]
  \begin{center}
 \mbox{\epsfig{figure=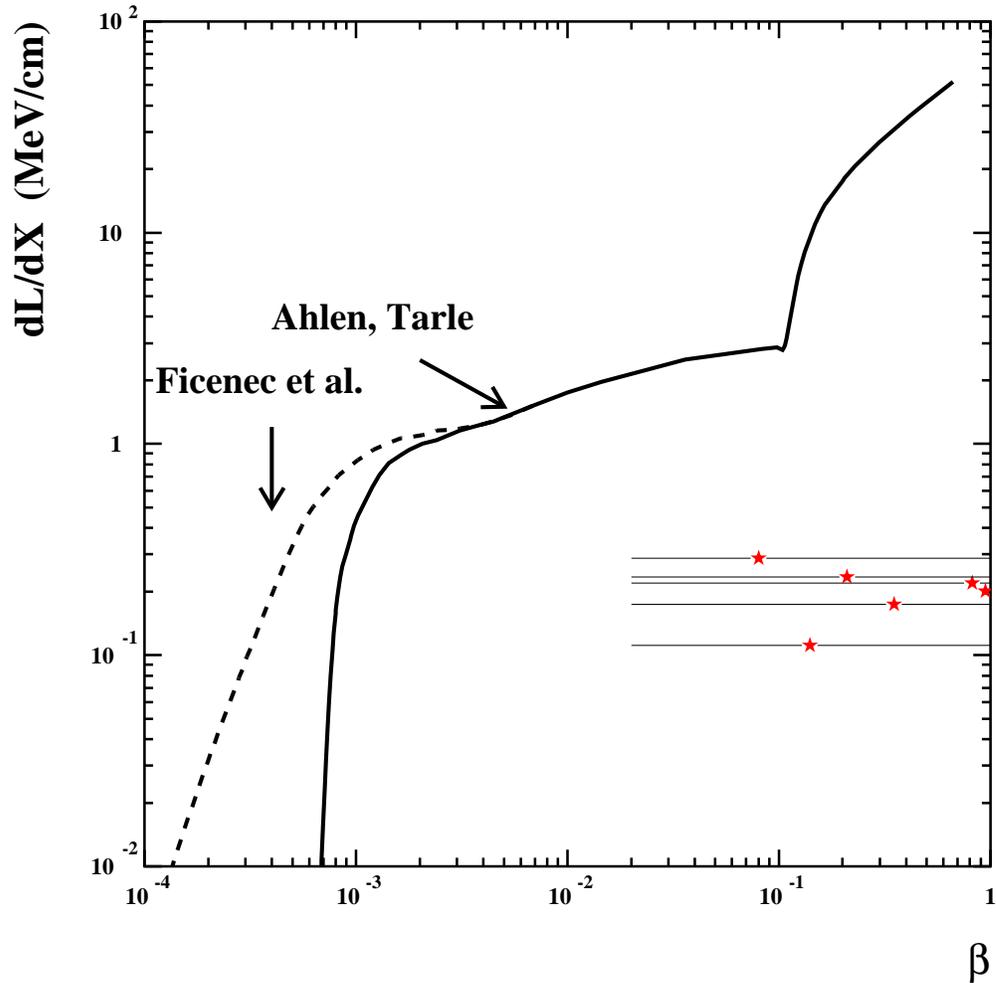,width=14cm,bbllx=31bp,bblly=200bp,bburx=550,bbury=630bp}}
   \vskip 0.5 cm
    \caption{\it \label{fig:dldxbeta}
             Light yield of the six selected candidates compared with the expected
             monopole signal as calculated in \protect\cite{ahlentarle,ficenec}.
             As can be seen for all the events the measured values are 
             well below the expectations for monopoles.}
  \end{center}
\end{figure}

As a further check, the measured values of energy loss (in the scintillator counters)
and $\beta$ of surviving candidates have been compared with
the expected monopole signal as a function of its velocity 
\cite{ahlentarle,ficenec}. This comparison is reported in 
Fig.\ref{fig:dldxbeta}. For five of the six events the Time of Flight (T.o.F.)
information was provided by the streamer system alone, since only one fired 
scintillator counter was present. In this case the error on 
the reconstructed velocity is large because of the limited time resolution 
of the streamer tubes.
As can be seen for all the events the measured energy loss values are 
well below the expectations for monopoles. The six events are interpreted 
as high energy muons producing energetic $\delta$ rays.

\subsection{Flux upper limit calculation}
\begin{figure}[hp]
  \begin{center}
 \mbox{\epsfig{figure=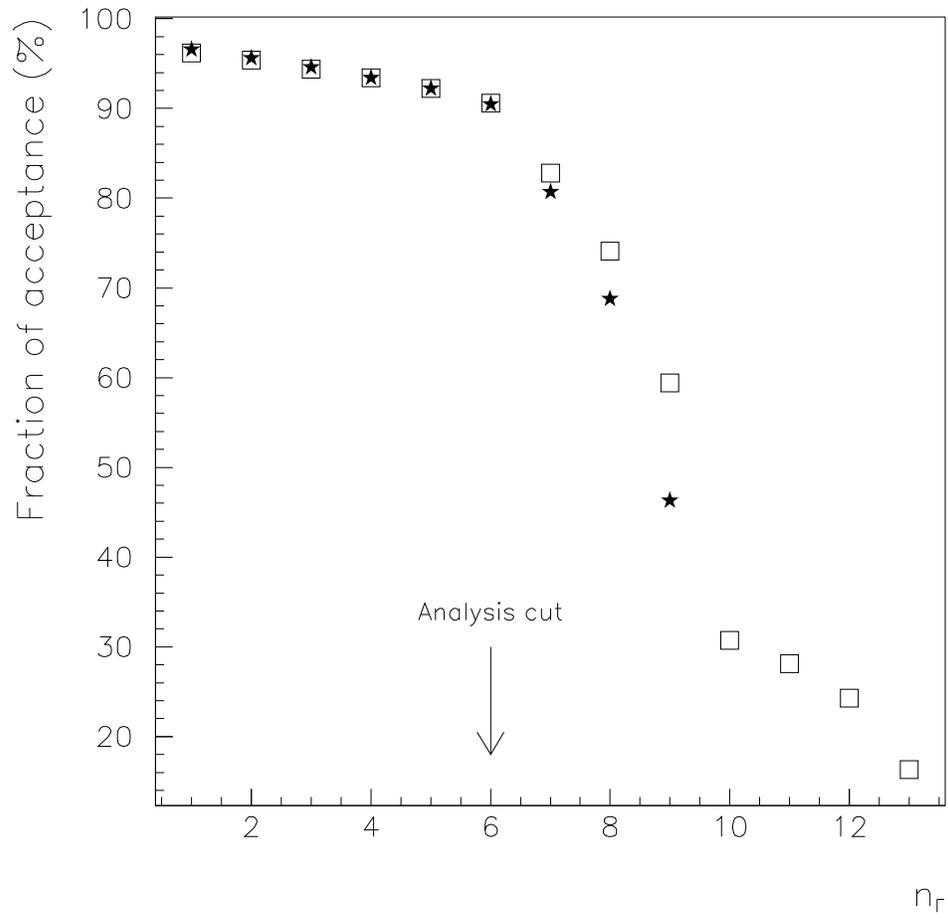,width=14cm,bbllx=31bp,bblly=200bp,bburx=550,bbury=630bp}}
   \vskip 1 cm
    \caption{\it \label{fig:perac}
             Fraction of the effective geometrical acceptance after all cuts
             as a function of the minimum required $n_{\Gamma}$.
             Stars refer to data simulated without considering the attico.
             The acceptance for  $n_{\Gamma} \ge 6$ is 3565$\, m^2 sr$.}
  \end{center}
\end{figure}

To evaluate the acceptance, 30$\,$000 events have been simulated and written
with the format of real data in order to use the same off-line code employed
for the analysis. Since in the first data set the upper part of
MACRO ({\it attico}) was not in operation, configurations both with and without 
it have been simulated. By taking into account all the cuts required by the
analysis, we find an acceptance of  $\simeq 3565 \, m^2 sr$.
In Fig.\ref{fig:perac} the acceptance dependence on the minimum imposed
$n_{\Gamma}$ is reported. In the analysis $n_{\Gamma} \geq 6$ is required.
By considering all the effects due to  electronics, triggers, cuts on 
$\Delta E$ and  $\Gamma$, etc., the product of the  efficiency times the live 
time integrated in the three periods is 11853 hours. 
Since no candidate was found, the monopole flux upper limit at $90 \%$ C.L. is
\begin{equation}
  \phi \le 1.5 \cdot 10^{-15} \, cm^{-2} s^{-1} sr^{-1}
\end{equation}
for $ 5 \cdot 10^{-3} \le \beta \le 0.99$ and monopole induced
nucleon decay catalysis cross section $\sigma \le 1\,$mb.
For $\beta \ge 0.99$ and $\sigma > 1\,$mb the efficiency of this search might be 
reduced by an amount which is difficult to estimate due to 
theoretical uncertainties (see Sec.\ref{sec:signa}).

This is the first search in the literature performed with three
different detection techniques. Furthermore,  in the velocity region 
$10^{-1} \leq \beta \leq 0.99$, the limit is more stringent than
any other obtained with scintillation or gas detectors  
(see for instance \cite{baksan,kgf,soudan2,imb}).

\section{Conclusions}
A combined analysis technique to search for fast magnetic monopoles
with the MACRO detector has been described.
The use of scintillator, streamer tubes, and track-etch data at the same time,
and the large energy release of fast monopoles,
provide a clear signature that allows us to reject the cosmic muon
background with highly reliable and safe cuts.
A first sample of more than two years of data has been analyzed.
No candidate survived thus setting a local monopole
flux  upper limit at 90$\%$ CL of
$ \phi \le 1.5 \cdot 10^{-15} \, cm^{-2} s^{-1} sr^{-1} $
for $5 \cdot 10^{-3} \le \beta \le 0.99$ and nucleon decay catalysis
cross section $\sigma \le 1 \, mb$.
Even with this reduced data sample in the region
$10^{-1} \leq \beta \leq 0.99$, the limit is more stringent than any other
obtained with scintillation or gas detectors.
By using all of the MACRO data (i.e. up to December 2000), the described analysis 
technique will be able to extend its sensitivity to fluxes of the order of 
$ 5 \cdot 10^{-16} \, cm^{-2} s^{-1} sr^{-1} $.

\vskip 1.cm
{\bf Acknowledgements. }\\
We gratefully acknowledge the support of the director and of the staff of the Laboratori 
Nazionali del Gran Sasso and the invaluable assistance of the technical staff of the 
Institutions participating in the experiment. We thank the Istituto Nazionale di Fisica 
Nucleare (INFN), the U.S. Department of Energy and the U.S. National Science Foundation 
for their generous support of the MACRO experiment. We thank INFN, ICTP (Trieste), 
WorldLab and NATO for providing fellowships and grants (FAI) for non Italian citizens.

\newpage 
 
%
%
 
\end{document}